\documentclass[runningheads]{llncs}
\usepackage[T1]{fontenc}
\usepackage[english]{babel}
\usepackage{enumitem}
\usepackage{graphicx}
\usepackage{glossaries}
\usepackage{pgfplots}
\usepackage{soul}
\usepackage{url}
\usepackage{rotating}
\usepackage{tablefootnote}
\usepackage{booktabs}
\usepackage{tabularx}
\usepackage{csquotes}
\usepackage{microtype}
\usepackage{cleveref}
\usepackage{siunitx}
\usepackage[framemethod=TikZ]{mdframed}

\pgfplotsset{compat=1.18}

\newmdenv[backgroundcolor=gray!20, roundcorner=5pt, linewidth=1pt, skipabove=1em]{rqanswer}

\begin{document}

\title{SoK: Towards Reproducibility for Software Packages in Scripting Language Ecosystems}

\author{
    Timo Pohl\inst{1}\orcidID{0009-0002-3760-7976} \and
    Pavel Novák\inst{2}\orcidID{0000-0002-3758-5757} \and
    Marc Ohm\inst{1,3}\orcidID{0000-0002-2913-5270} \and
    Michael Meier\inst{1,3}\orcidID{0009-0006-8199-5004}
}
\authorrunning{T. Pohl et al.}
\institute{
    University of Bonn, Germany \\
    \email{\{pohl,ohm,mm\}@cs.uni-bonn.de} \\
    \and
    Masaryk University, Czech Republic \\
    \email{468940@muni.cz} \\
    \and
    Fraunhofer FKIE, Germany
}

\maketitle              

\begin{abstract}
    The disconnect between distributed software artifacts and their supposed source code enables attackers to leverage the build process for inserting malicious functionality.
    Past research in this field focuses on compiled language ecosystems, mostly analysing Linux distribution packages.
    However, the popular scripting language ecosystems potentially face unique issues given the systematic difference in distributed artifacts.
    This SoK provides an overview of existing research, aiming to highlight future directions, as well as chances to transfer existing knowledge from compiled language ecosystems.
    To that end, we work out key aspects in current research, systematize identified challenges for software reproducibility, and map them between the ecosystems.
    We find that the literature is sparse, focusing on few individual problems and ecosystems.
    This allows us to effectively identify next steps to improve reproducibility in this field.
\end{abstract}

\section{Introduction}

Many programming ecosystems encourage the use of so-called \emph{software packages} to improve development speed and reduce redundant work.
These software packages, often libraries or frameworks, usually perform common tasks that can be generalized in a way that they can be used in a broad range of contexts, leading to some packages being part of hundreds of thousands of software projects.
This of course makes them attractive for attackers, as infecting a single one of those packages could potentially infect all the machines running any of the dependent software.
Such attacks are called software supply chain attacks, since they abuse the dependency relation between packages to indirectly infect a dependent package.

With ongoing research on detecting malicious snippets in source code~\cite{gonzalez2021anomalicious,ohm2022feasibility,sejfia2022practical}, other attack vectors are explored to infect software packages.
One of them is the infection of the build process.
Software packages distribution usually happens via so-called \emph{package registries}, where artifacts can be uploaded without any verifiable link to the source code they are supposedly built from.
Therefore, even if the alleged source code of a package is carefully vetted, the distributed artifact could perform completely different tasks.
There have been recent incidents exploiting exactly this circumstance, like the \texttt{ultralytics} Python package that was hit by a cache-poisoning attack on the GitHub Actions build process, leading to an artifact published on PyPI that contained functionality which was not present in the GitHub source code it was built from~\cite{socket2024ultralytics}.

A strategy to combat this is to build the artifact at multiple independent builders and compare their artifacts.
Ideally, all the artifacts would be the same when built from the published source code, suggesting that no builder has added any malicious changes.
However, it has been shown that even if all parties are benign, differences occur across different builders~\cite{bajaj2023unreproducible,benedetti2025empirical,drexel2024reproducible}.
This happens because the tools used for building the artifacts are not optimized for reproducibility, and often generate non-deterministic output, like a timestamp of the build time.
Thus, to enable this strategy of combatting malicious or compromised build services, these build processes have to be made reproducible.

According to the StackOverflow Developer Survey 2024~\cite{stackoverflow2024survey}, the most used programming languages are JavaScript (62.3~\%), Python (51~\%) and TypeScript (38.5~\%), all of which are scripting languages.
While compiled languages, especially in the context of Linux distribution packages for Debian and Arch Linux, have received a lot of attention regarding problems and solutions to achieve reproducibility~\cite{fourne2023it}, scripting language ecosystems for languages like JavaScript or Python, are not reproducible for the most part.
This paper systematizes and collects the current research results regarding this topic, aiming to identify well researched areas, fields in which knowledge may be transferred from compiled language ecosystems, as well as future research directions.
In particular, we will answer the following research questions:
\begin{enumerate}[label=\bfseries RQ\arabic*\normalfont]
    \item What are the key aspects of reproducibility research for software packages in scripting language ecosystems?
    \item What are common challenges for reproducibility in compiled and scripting language ecosystems?
    \item What are \emph{additional} challenges for reproducibility in scripting language ecosystems?
\end{enumerate}

The remainder of the paper is structured as follows:
\Cref{sec:background} provides background on reproducible builds, focusing on the definition, high-level approaches and the main differences concerning compiled and scripting language ecosystems.
\Cref{sec:state_of_the_art} then provides an overview of the current state of the art in reproducibility\footnote{For brevity, we use the generic term \enquote{reproducibility} to mean \enquote{software reproducibility}.} research with respect to scripting language ecosystems, answering \emph{RQ1}.
\Cref{sec:common_challenges} compiles and compares challenges regarding reproducibility in compiled and scripting language ecosystems, answering \emph{RQ2} and \emph{RQ3}.
In \Cref{sec:discussion} we discuss our findings, followed by an outlook on future research directions in \Cref{sec:future-research-directions}.
The paper is concluded in \Cref{sec:conclusion}.

\section{Background}\label{sec:background}

In order to better understand our motivation and the contents of this paper, this section provides background on software reproducibility.
We focus on its goals and their relation to different definitions found in the literature and the industry, as well as background on the differences between compiled and scripting language ecosystems in this context.

\subsection{Software Reproducibility}

The exact definition of software reproducibility varies across the literature and the industry.
The main goals of software reproducibility, as stated by the Reproducible Builds project~\cite{reproducibleBuildsProject}, are security and quality assurance.
Regarding security, it aims to achieve resistance against attacks in the build process, which includes all the steps performed to create an artifact from given source code.
In terms of quality assurance, it is supposed to guarantee that binaries behave the same, no matter on what system they were built.

The general idea behind reproducibility is always the same:
Given the same input, the executable artifact resulting from the build process shall always behave the same.
However, details in the definitions exhibit differences in the following three factors:
\begin{itemize}
    \item How is behavioural equivalence tested or approximated?
    \item What exactly is part of the input?
    \item What is the build process?
\end{itemize}

Lakhotia et al.\ have shown that, in general, deciding whether two binaries behave equivalently is impossible~\cite{lakhotia2013fast}.
For this reason, Dietrich et al.\ have formalized a set of criteria for equivalence relations between binaries, resulting in four assurance levels for behavioural equivalence~\cite{dietrich2024levels}.

\emph{Level 1} equivalence relations simply compare for bit-by-bit equivalence.
\emph{Level 2} equivalence relations perform bit-by-bit comparisons too but only in parts of the binary that are regarded as having a semantic effect on the binaries' behaviour.
This excludes parts like embedded signatures, only used to check the authenticity of the binary by a third party.
\emph{Level 3} equivalence relations do not require any bit-by-bit comparisons.
They regard semantically equivalent code sequences as equivalent.
\emph{Level 4} equivalence relations regard semantically \emph{similar} code as equivalent.~\cite{dietrich2024levels}

Level 1 equivalence relations are easy to perform and check.
However, it quickly happens that two binaries that behave the same are not regarded as equivalent by these relations.
Level 2 and level 3 relations are less prone to falsely label equivalent behaviour as non-equivalent.
Their drawback is that it is much harder to create relations that can reliably identify parts of artifacts with a semantic effect, or that are semantically equivalent than a simple bit-by-bit comparison, especially if these relations have to be generic over a heterogeneous set of artifacts.
Level 4 equivalence relations do not actually test for behavioural equivalence but rather for behavioural similarity, thus being unable to achieve the goals stated above for reproducible software.

The most commonly used equivalence relation regarding reproducible builds is the level 1 bit-by-bit equivalence of the whole build artifact, used by almost all industry projects.
Carnavalet et al.~\cite{carnavalet2014challenges} introduce the notion of \emph{verifiable builds} which allow omitting \enquote{unimportant details} from the comparison, corresponding to a level 2 equivalence relation.
The Fedora Project~\cite{fedoraReproducibleBuilds} follow this notion, though still calling it reproducible builds, and ignore the embedded signatures of their software packages when comparing them.
Pöll et al.\ also propose the notion of \enquote{accountable builds}, ignoring all differences that are explainable and can be considered benign upon inspection~\cite{poll2022automating}, which would correspond to level 4 equivalence.
We see no industry projects following this definition.

Regarding the definition of inputs, there are different granularities in how exact the input is defined.
The Yocto Project~\cite{yoctoReproducibleBuilds} requires \enquote{build configurations} to be the same, while VMWare~\cite{vmwareReproducibleBuilds} says that \enquote{all states of the build} have to be controlled.
The Reproducible Builds Project~\cite{reproducibleBuildsProject} defines the input as the \enquote{same source code, build environment and build instructions}.
The Fedora project also adds \enquote{metadata} to these required inputs.

Some projects also explicitly state what the reproducibility should \emph{not} depend on.
FOSSA Inc.~\cite{fossaReproducibleBuilds} says that the build should not depend on the computer, the time or the available network services.
The Yocto Project~\cite{yoctoReproducibleBuilds} states that builds should be reproducible regardless of the system they are run on.

Even though some of these definitions explicitly name inputs like \enquote{build configuration} or \enquote{build instructions}, what exactly is part of them is not clearly defined.
The Reproducible Builds project even explicitly states that maintainers have to choose the \emph{relevant} parts of the build environment, and that these relevant parts should be as small as possible.~\cite{reproducibleBuildsProject}.
Thus, under all of these definitions it is still up to maintainers what exact parts of these inputs they provide.

At last, the definitions of what exactly is meant with the build process is mostly a concern of software packages in scripting languages.
In compiled languages the build process consists of the compilation of the source code to an executable binary or bytecode, including all necessary pre- and post-processing.
In scripting languages, some research considers the installation of a software package as part of a \enquote{reproducible build}~\cite{mukherjee2021fixing}, while other research considers the actual process of creating a package that can be distributed as an individual file to users of the software from the source code as the build process~\cite{benedetti2025empirical,goswami2020investigating}.
The relevant difference is that the latter only depends on \emph{build dependencies}, while the former depends on \emph{runtime dependencies}.
In the context of scripting language ecosystems, the generic term \emph{dependencies} usually refers to \emph{runtime dependencies}, while in compiled language ecosystems, it often refers to \emph{build dependencies}.

For this paper, we arrive at \Cref{def:rep} which is a more formal version of the definition given by the Reproducible Builds project.
In particular, it explicitly states that reproducibility is a property of a tuple (source code, build instructions, build environment, artifacts).
Throughout our paper, we consider tuples fulfilling this definition to have the \emph{reproducible builds} property.

\begin{definition}\label{def:rep}
    A tuple (source code, build instructions, build environment, artifacts) is considered reproducible, if executing the build instructions on the source code within the build environment always produces the same artifacts when compared via bit-by-bit equality.
\end{definition}

However, proving reproducibility is difficult, as artifact equality would have to be proven under \emph{all} circumstances.
Therefore, the common practice to consider such a tuple as reproducible is what Lamb and Zacchiroli have coined \enquote{adversarial rebuilding}~\cite{lamb2022reproducible} \textemdash{} deliberately varying external factors, especially those known to cause irreproducibility, and checking whether the artifact stays the same.
This can be achieved with tools like \texttt{reprotest}\footnote{\url{https://salsa.debian.org/reproducible-builds/reprotest}}.

While the \emph{reproducible builds} property is useful when trying to verify that a certain artifact was not illegitimately modified during the build process, it is neither strictly necessary nor sufficient, especially in ecosystems that do not (yet) aim to achieve build reproducibility.
For this task, which we will call \emph{artifact reproduction}, a software artifact is given as input and the goal is to reproduce this given artifact.
This is commonly done for artifacts as they are distributed via registries.
In ecosystems where it is not yet common to be concerned with build reproducibility, research in artifact reproduction may help to establish independent verifiers for distributed packages.
This task can pose additional challenges, for example acquiring source code for an artifact, which reproducible builds assumes to be given.

\subsection{Achieving Reproducibility}

Lamb and Zacchiroli~\cite{lamb2022reproducible} split the reasons for irreproducibility in two broad categories: unstable inputs \textemdash{} for example build dependencies that are downloaded over the network during the build process, thus possibly returning different data in two distinct build runs \textemdash{} and non-determinism of the build tools, like compilers not always translating the same source code to the same machine code.

On a high level, the approaches to achieve build reproducibility can be split into proactive and reactive.
The proactive approach involves controlling and stabilizing all inputs to the build process, as well as ensuring that all build tools are fully deterministic.
While this approach should generically work for all build processes, guaranteeing access to all dependencies and the determinism of all involved build tools can be a demanding task.
It typically involves containerization, dependency version pinning, and well-defined build scripts~\cite{navarro2020reproducible}.

In the reactive approach, parts of the artifact known to be affected by non-determinism are patched after the build to always contain the same content.
While this method does not require control over external factors, it needs a deep understanding of the concrete build process, to identify parts of the artifact affected by non-determinism and patch them without breaking the functionality of the program.
Additionally, this method cannot resolve dependency-related issues~\cite{ren2022automated,randrianaina2024options}.
Debian's \texttt{strip-nondeterminism}\footnote{\url{https://salsa.debian.org/reproducible-builds/strip-nondeterminism}} is an example of a tool to perform such post-processing.

\subsection{Artifacts in Compiled- and Scripting Language Ecosystems}

On a high level the build process in compiled and scripting language ecosystems is the same.
A set of build instructions is executed on the source code, transforming it into one or few build artifacts.
However, there are differences between artifacts in compiled and scripting language ecosystems that require dedicated research.

The main difference is that while build artifacts in compiled languages typically consist of machine or bytecode, artifacts for scripting languages are customarily some form of archive consisting of source code files.
Additionally, ecosystems like NodeJS also commonly involve heavy preprocessing like transpilation, minification or bundling, posing unique challenges for reproducibility.
Furthermore, archive formats often contain metadata both about themselves and the contained files.

Yet, many scripting language ecosystems also support so-called \emph{native extensions}, meaning that some functionality is externalized into compiled binaries, often for performance reasons.
Similarly, some ecosystems like Python also support pre-compiling parts of the source code into bytecode, which can be distributed in packages for performance benefits.
As these binaries also have to be compiled from source code at some point, they may suffer from the same irreproducibility issues as software projects in compiled language ecosystems in general.
We thus argue that the problems one may encounter in scripting language ecosystems is a superset of the problems encountered in compiled language ecosystems.

\section{State of the Art in Reproducibility Research for Scripting Language Ecosystems}\label{sec:state_of_the_art}

This section answers \emph{RQ1} through a literature review of research investigating reproducibility in scripting language ecosystems.
We start with reproducible builds, highlighting that discrepancies between artifacts exist, as well as categorizing these discrepancies.
Afterwards research on artifact reproduction is investigated, mainly dealing with categorizing existing issues, as well as discovering the source code for a given artifact.
At last, we show the development within popular scripting language ecosystems that assist build reproducibility.

Vu et al.~\cite{vu2021lastpymile} have investigated the differences between Python packages distributed on the PyPI registry and the corresponding source code.
From the \num{4000} most downloaded packages, they selected all packages with working, unique links to GitHub repositories, which contain a complete history of commits for the whole project lifetime.
However, how exactly this commit completeness was evaluated is not clearly stated.
This resulted in a sample of \num{2438} packages, and \num{93252} artifacts considering all their versions.

They show that \qty{65}{\percent} of these artifacts exhibit differences compared to the linked source code repository.
While this is not a direct violation of the reproducible builds property, it indicates that some preprocessing steps are performed before packaging the source code files.
Regarding reproducibility verification, it also shows that reliably finding the source code that was used to create a package is not as easy as comparing the individual files of the software package to the files in the source code repository, and assuming that they should be equal.

Benedetti et al.~\cite{benedetti2025empirical} investigated reproducibility for software components across different ecosystems.
The scripting language ecosystems they investigated are NodeJS, Python and Ruby.
For these scripting languages, they investigated the number of reproducible packages, potential root causes and the influence of native extensions (c.f.~\Cref{sec:background}).

They randomly sampled \num{4000} packages from the ecosyste.ms\footnote{\url{https://ecosyste.ms}} project, for packages fulfilling the following criteria:
Contains the respective ecosystem's metadata file like a \texttt{package.json} for NodeJS packages, has a working link to a source code repository, and a successfully terminating build process.
Thus, general build failures would not count towards root causes for irreproducibility.

Using the tool \texttt{reprotest} to alter environmental conditions like the system time, build path and more, they built an artifact out of the most recent version of the repository code multiple times, and compared the results.
They consider the immediate output artifacts of the native build commands as the artifact to compare, with no special post-processing like unpacking tarballs.
A package is labelled reproducible, if all runs for all environmental variations with \texttt{reprotest} result in the same artifact.~\cite{benedetti2025empirical}

They show that \qty{100}{\percent} of their tested NodeJS packages, \qty{12.2}{\percent} of Python and \qty{0}{\percent} of Ruby packages are reproducible without any modifications.
Manually setting up the environment for reproducibility, for example by setting the \texttt{SOURCE\_DATE\_EPOCH} environment variable\footnote{\url{https://reproducible-builds.org/specs/source-date-epoch/}} increased reproducibility of Ruby packages to \qty{97.1}{\percent}, while not having an effect on Python package reproducibility.
Additionally, patching the package manager \texttt{pip} to set predictable file permission bits increased reproducibility of PyPI packages to \qty{98}{\percent}.
They also found that timestamps were responsible for \qty{87.77}{\percent} of Python, and \qty{97.1}{\percent} of unreproducible Ruby packages.~\cite{benedetti2025empirical}

For the rest of unreproducible builds, they mainly blamed the ability to run arbitrary code during the packing process, expressing that fixing irreproducibility due to arbitrary code execution is hard to combat by nature~\cite{benedetti2025empirical}.
While this is only explicitly mentioned for the Python and Ruby ecosystems, we note that \texttt{npm} has the same ability with their lifecycle scripts for the \texttt{pack} event, that would be automatically executed when calling \texttt{npm pack}~\cite{npmScriptsDocs}.
However, as we will see later this section in the work of Goswami et al.~\cite{goswami2020investigating}, it has become common practice in the NodeJS ecosystem to use the \texttt{build} script instead of these lifecycle scripts.
We assume that this is the reason why they do not affect the NodeJS ecosystem much in practice.

While this is valuable research, we argue that this tests the native build process of the ecosystem itself, rather than the reproducibility of individual packages.
For example, the previously mentioned \texttt{build} scripts commonly employed in the NodeJS ecosystem are not automatically executed by the native build command \texttt{npm pack}.
Since preprocessing with transpilers, bundlers, minifiers and other tools is common practice, an experiment testing the package managers' build commands in isolation does not give insight into the reproducibility of the whole process from published source code to distributed artifact.
Still, testing the package managers in isolation is a valuable step in identifying parts of the process that lead to irreproducibility.

Regarding the influence of native extensions to reproducibility, Benedetti et al.\ find that native extensions do not change the distribution of irreproducible packages, even though the C and C++ compilation process has shown to have many issues regarding reproducibility, if not explicitly addressed~\cite{bajaj2023unreproducible,drexel2024reproducible,fourne2023it,lamb2022reproducible,poll2022automating,ren2019root,tomassi2019bugswarm,carnavalet2014challenges}.
For NodeJS and Python this is attributed to their native extension toolchains, which are built with reproducibility in mind.
However, Ruby does not have such a toolchain.
They conjecture that native extensions are often a collection of small independent functions for very specific, computationally expensive workloads, and that this leads to little influence of typical irreproducibility causes.~\cite{benedetti2025empirical}

In face of the little research regarding reproducible builds for interpreted language ecosystems, there is more work regarding artifact reproduction in this context.
Goswami et al.~\cite{goswami2020investigating} have investigated how many of the \num{1000} most depended upon packages in the npm registry they are able to reproduce, and established key challenges in doing so.
To find the matching source code, they used the repository link given in the package metadata, if available.
As all the available repositories pointed to GitHub, they used the GitHub releases to identify commits that presumably correspond to released artifacts on the npm registry.
For all releases found on GitHub, they matched the release's name to a version released on npm.
This way, they established a set of pairs of source code, and a corresponding distributed artifact.
Afterwards, all packages that did not contain a \texttt{package.json} file, which is used to store package metadata, as well as all packages that do not contain a \texttt{build} script within the \texttt{package.json} were removed.
Thus, the resulting dataset contained a total of \num{3390} versions for \num{226} packages.

To build the packages from the source code, they ran the commands \texttt{npm install} and \texttt{npm run build}~\cite{goswami2020investigating}.
The former command installs all the defined runtime dependencies, as well as development dependencies~\cite{npmInstallDocs}, while the latter runs an optional, user-defined script named \texttt{build}.

While the name might wrongfully suggest that this is a semantically meaningful script which is somehow standardized or encouraged to be used for the package build process, this is not the case.~\cite{npmScriptsDocs}.
This procedure also does not actually run the \texttt{npm pack} or \texttt{npm publish} commands, which are the commands to actually create the tarball artifact and publish it to the npm registry respectively~\cite{npmCLIdocs}.
Thus, they also skip all the lifecycle scripts a package maintainer might have defined for these operations.
This also means that their resulting artifact used for later analysis is not the actual tarball that is distributed by the npm registry, but only its contents.

While it is not explicitly mentioned, we assume that they unpack the registry artifact for comparison, in order to at least have artifacts of the same form to compare.
They use \emph{diffoscope}\footnote{\url{https://diffoscope.org}} to examine the difference between their built artifact, and the corresponding registry artifact~\cite{goswami2020investigating}.

They find that almost \qty{40}{\percent} of package versions are not reproducible.
For \num{492} packages (\qty{15}{\percent}) the build process failed.
For the remaining \num{2898} package versions, they investigated and categorized the differences in source code files, and analysed potential root causes, which are further discussed in \Cref{sec:common_challenges}.~\cite{goswami2020investigating}

In 2021 Vu proposed \emph{py2src} to investigate how to link Python packages to their corresponding source code~\cite{vu2021py2src}.
This tool takes various information sources that may include a link to a source code repository into account, like information returned by the PyPI registry itself, or information from metadata files within the package, and rates them by how likely they are the correct link.
They propose six heuristics to judge a single URLs reliability: Similarity of package and repository name, similarity of GitHub repository and PyPI package descriptions, \enquote{Python} being part of the GitHub repositories' reported languages, GitHub page containing badge that links to PyPI package, PyPI maintainer name being present in GitHub contributor list and GitHub repository tags matching versions released on PyPI.

These heuristic indicate that both, the GitHub repository and the Python package are actually maintained by the same author.
However, they do not help to establish a trusted link between the artifact and the source code in case of a malicious author, nor do they contribute in finding the correct source code revision for a given version one may want to rebuild.
Their reliance on package metadata also does not help to find source code for packages that do not link to a source code repository in their metadata.

Tsakpinis and Pretschner~\cite{tsakpinis2024analyzing} picked up this last problem, and examined the availability of source code for the Python and NodeJS ecosystems.
They collected all metadata for all packages in the PyPI and npm registries respectively, and analysed the values in the respective source URL metadata fields.
If the given URL pointed to a repository on GitHub, they also analysed whether it was publicly available.
They found that \qty{42}{\percent} of Python packages, and \qty{54}{\percent} of NodeJS packages have links that do not point to a valid GitHub repository, with \qty{24}{\percent} and \qty{50}{\percent} respectively having known invalid or missing links, further highlighting the problem of relying on package metadata to find the corresponding source code

Gao et al.~\cite{gao2024pyradar} propose \emph{PyRadar} aiming to find the corresponding source code repository to a Python package based on the code of the artifact, as well as validating a supposed link between a source code repository and a given artifact.
Regarding the repository validation, they leverage so-called \emph{phantom files} \textemdash{} files that appear in the registry artifact but not in the corresponding source code repository \textemdash{} as a metric.
A file is considered a phantom file if the source code repository does not contain a file with the same SHA-256 hash.~\cite{gao2024pyradar}

To evaluate their approach, they take about \num{14000} Python packages that are present on GitHub with at least \num{100} stars, for which they can find a corresponding package on PyPI, as the dataset with assumed correct links between the source code repository and registry artifact.
To gather packages with incorrect links, they sample all Python packages that point to a repository already present in the presumed correct links dataset, but with a different listed maintainer.
Furthermore, they add all Python packages that list the default \texttt{pypa/sampleproject} GitHub repository as their source code repository.
This way, they end up with about \num{2000} packages in their presumed incorrect links dataset.
These datasets are manually verified for a \qty{95}{\percent} confidence level and \qty{5}{\percent} confidence interval.~\cite{gao2024pyradar}

They show that there is a significant difference in occurrence of phantom files between artifacts and the source code from the correct source code repository compared to the source code from a different source code repository.
They also show that certain files, like the Python package metadata files \texttt{setup.py} and \texttt{pyproject.toml} have a particularly high likelihood to be phantom files when comparing against incorrect source code repositories, while they are present within the correct source code repositories.~\cite{gao2024pyradar}

Including these findings, they come up with six features to train machine learning models on that can be used for classification whether a given link points to the correct source code repository or not.
These are the number of phantom Python files, whether one of the Python package metadata files is a phantom file, whether the version of the package in question has a matching tag in the repository, the normalized Levenshtein similarity between the Python package name and the repository name, the number of maintainers of the package and the number of packages maintained by the package maintainers.
Out of seven models they evaluated, a Random Forest model performs best with an \emph{area under the ROC curve} of 0.995.~\cite{gao2024pyradar}

While this is a promising approach to verify the correctness of the source code repository links, this is not a safe method to verify that a given registry artifact actually stems from the source code in the linked repository, but rather a plausibility check that the repository has not been accidentally mislinked.

Regarding source code retrieval for packages with no (correct) link to a source code repository, they leverage the \emph{World of Code}\footnote{\url{https://worldofcode.org/}} infrastructure to perform hash-based git repository queries.
For a given Python artifact, they iterate over all Python files and calculate their SHA-256 hashes.
They use these hash to look up all the commits the respective file is introduced in.
Afterwards, these commits are mapped to unique repositories they appear in.
If the file appears in less than a configurable amount of repositories, all those repositories are added to a set of candidate repositories for the artifact.
Candidate repositories are then ranked by the amount of files from the artifact they contain.
At last, the name similarity of the Python package and the top ranked source code repository name is calculated.
If it is below a configurable threshold, no repository is returned.
Otherwise, the top ranked source code repository is returned.~\cite{gao2024pyradar}

Using heuristically set values of \num{500} for the maximum amount of repositories a file may appear in, and \num{0.5} for the minimum name similarity between the package and the repository, \qty{90.2}{\percent} of repositories from their correct links dataset return a repository candidate, with an accuracy of \num{0.97}.~\cite{gao2024pyradar}

This is a promising approach to find source code repositories for python packages.
It leverages the fact that in python the python ecosystem, individual files often appear unaltered in the resulting package.
In other ecosystems like NodeJS, this is not necessarily the case, due to its common use of tools like minifiers or transpilers, transforming the source files before publishing them into registry packages.
The transferability of this approach to other ecosystems is therefore to be evaluated.

Aside from these academic achievements, we have also observed efforts to work towards reproducible builds from within ecosystems.
PEP 552~\cite{pep552} was finalized in 2017, which outlines techniques to make compilation of python bytecode files more deterministic.
Python core developers also created a tool called \textit{Asaman}\footnote{\url{https://github.com/kushaldas/asaman}}, aiming to produce reproducible wheels.
However, according to its own documentation, this tool still has many limitations, and at the time of writing the latest commit was almost two years ago in April 2023, indicating that it may no longer be actively maintained.

Recently, two package registries also started to support \emph{digital attestations of provenance information}.
Namely, npm in early 2023~\cite{npmPackageProvenance} and PyPI in late 2024~\cite{pypiPackageProvenance,pep710} introduced the ability to attach signed attestations to package releases, attesting that they were built from a certain source code revision with a specified build script.
When these are signed by trusted entities, which could be for example the GitHub or GitLab organizations, they provide a trustworthy link between a package release and the corresponding source code, including the exact source code revision, build scripts and the build environment.
If widely adopted, this could be very beneficial for artifact reproduction, as it eliminates the need to find the exact source code, build instructions and build environment a given artifact was built in.

In summary, we see that there is little research regarding reproducible builds in scripting language ecosystems, covering multiple ecosystems but investigating the isolated influence of package managers and their native build commands to reproducible builds.
For artifact reproduction, there is a little more research, but individual topics often focus on a single ecosystem.
Systematic analysis of challenges for artifact reproduction has only been done for the NodeJS ecosystem, while research into discovering the source code for a given artifact is focused on the Python ecosystem.
Overall, research in this context is very limited, and therefore hard to generalize over scripting language ecosystems as a whole.

\begin{rqanswer}
    \textbf{Response to RQ1:}
    Research regarding reproducibility for software packages in scripting language ecosystems focuses on artifact reproduction.
    Two common subtopics are root cause analysis for irreproducibility, as well as identification of the source code for a given artifact.
    A single recent publication investigates the effect of package managers on reproducible builds.
\end{rqanswer}

\section{Challenges for Reproducible Builds in Compiled- and Scripting Language Ecosystems}\label{sec:common_challenges}

After establishing the key aspects of reproducibility research in scripting language ecosystems, this section collects challenges for the more researched compiled language ecosystems, as well as for scripting language ecosystems.
We aim to identify an overlap between the two, as well as additional challenges that scripting language ecosystems might be facing.
To do so, we briefly summarize known challenges for reproducible builds, first in compiled and then in scripting language ecosystems.
We then lay out which challenges they have in common, possibly with existing solutions, as well as challenges unique to scripting language ecosystems.

\subsection{Reasons for Irreproducibility in Compiled Language Ecosystems}

Regarding compiled language ecosystems, we first collect challenges identified for reproducible builds, following up with challenges for artifact reproduction.

\subsubsection{Reproducible Builds}

\begin{figure}
    \begin{center}
        \includegraphics{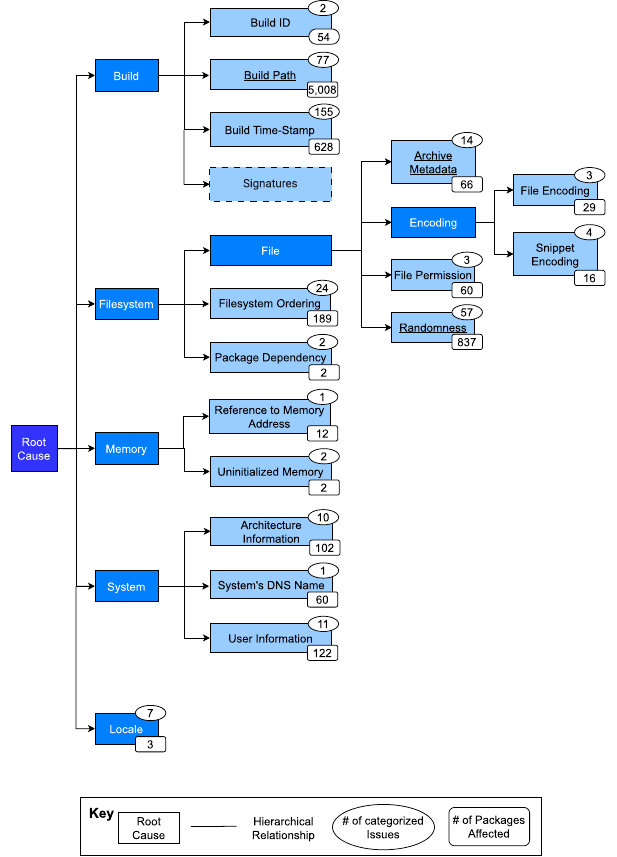}
    \end{center}
    \caption{
    Taxonomy of root causes for irreproducible builds in compiled language ecosystems by Bajaj et al.~\cite{bajaj2023unreproducible}.
    The number of issues belonging to a certain category is noted in the round box on the top right, with the number of affected packages noted in the rectangular box on the bottom right.
    The dashed box labelled \enquote{Signatures} has been amended by us as an additional root cause that has come up in other related work.
    }\label{fig:taxonomy}
\end{figure}

The reasons for build irreproducibility in compiled language ecosystems have been extensively researched.
In 2019 Ren et al.~\cite{ren2019root} used system call tracing during the build process of \num{180} randomly chosen Debian packages, identifying five main categories causing irreproducibility: Timestamps, randomness, file ordering, locale and user- or hostnames in the artifact.
A 2022 report by Lamb and Zacchiroli~\cite{lamb2022reproducible} also mentions build paths, archive metadata and uninitialized memory as main reasons for irreproducibility.
In 2023, Bajaj et al.~\cite{bajaj2023unreproducible} created a comprehensive taxonomy of irreproducibility causes, as well as case studies on their frequency and impact on packages for the Linux distributions Arch Linux and Debian.
Their taxonomy is displayed in \Cref{fig:taxonomy}.
The Reproducible Builds project~\cite{reproducibleBuildsProject} and Debian~\cite{debianIrreproducibilityReasons} maintain continuously updated lists of causes for irreproducibility, matching this taxonomy.

An interview with participants of the Reproducible Builds project performed by Fourné et al.~\cite{fourne2023it} has confirmed that the highest impact root causes of timestamps, build paths and general randomness are also perceived as the most pressing obstacles when trying to achieve reproducible builds.
They additionally bring up signatures~\cite{fourne2023it}, which are sometimes part of the build artifacts.
They do not appear in the previously mentioned taxonomy, because neither Arch Linux nor Debian include signatures as part of the package itself but rather as a separate detached artifact.
However, other Linux distributions like Fedora do embed signatures into their packages~\cite{fedoraReproducibleBuilds}.
We have therefore amended \Cref{fig:taxonomy} to also include signatures as a root cause.

\subsubsection{Artifact Reproduction}

Artifact reproduction in compiled language ecosystems is not researched as well, with only three papers explicitly looking into the topic.
In 2014 Carnavalet et al.~\cite{carnavalet2014challenges} performed a case study, trying to reproduce the 16 most recent versions of the disk encryption software \emph{TrueCrypt}\footnote{\url{https://www.truecrypt.org/}}.
Aside from reproducibility issues shown in the previously mentioned taxonomy displayed in \Cref{fig:taxonomy}, they had problems getting access to build tools.
On the one hand, not all the build tools used by the original developers are free to use.
On the other hand, older versions of the software required older, compatible versions of some build tools to build at all.

In 2024 Pöll and Roland~\cite{poll2022automating} analysed the reproducibility of android builds at the example of so-called \emph{Generic System Images}.
In addition to the previously mentioned reproducibility problems, they also faced the challenge of matching the correct source code to the artifact.
While they based their own builds on git tags matching the version of the image they want to rebuild, later analysis revealed that the respective Generic System Image was built from a later commit without a version tag.

Keshani et al.~\cite{keshani2024aroma} tried reproducing Java packages from the Apache Maven registry.
The main problems they experienced were missing links to source code repositories for given packages, missing tags denoting the exact revision of the source code that the package was built on, tags that by name match the version they try to rebuild but not actually point to the correct revision, and missing build environment definitions.

In summary, in addition to the challenges that general reproducible builds are facing, identified challenges unique to artifact reproduction are build tool availability and source code matching.
While the low amount of studies combined the nature of these rather small case studies does not allow concluding that this is an exhaustive list of challenges regarding artifact reproduction in compiled language ecosystems, is serves our aim to map already identified challenges in this field to the field of scripting language ecosystems.

\subsection{Reasons for Irreproducibility in Scripting Language Ecosystems}\label{sec:main_reasons_of_irreproducibility}

While fewer in number than for compiled language ecosystems, some studies also analysed reproducibility in scripting language ecosystems, for both reproducible builds and artifact reproduction.
All the mentioned studies are described in detail in \Cref{sec:state_of_the_art}.
This section highlights the mentioned reasons for irreproducibility.

\subsubsection{Reproducible Builds}

The only study to date that has performed an actual analysis of reproducible builds in scripting language ecosystems was performed by Benedetti et al.~\cite{benedetti2025empirical}.
For the scripting language ecosystems surrounding Ruby, Python and NodeJS, they investigated how reproducible the native build pipeline of the ecosystem is, without external pre- or post-processing.
They found that timestamps and file permissions were the largest contributors to irreproducibility.

Some of these problems can be solved by configuration.
For example, using the previously mentioned \texttt{SOURCE\_DATE\_EPOCH} environment variable is supported by \texttt{pip}.
Likewise, \texttt{pip} can be configured to use different backends that actually perform the package builds, of which some deterministically fix the permissions for the packaged files, while others do not.
However, none of the package managers under test have documentation indicating how to best configure them in order to achieve reproducible package builds.~\cite{benedetti2025empirical}

As mentioned in \Cref{sec:state_of_the_art} they also explicitly investigated how much native extensions \textemdash{} parts of the package written in a compiled language like C and called via foreign function interface \textemdash{} contribute to irreproducibility of these packages.
While one might expect that these native extensions fall victim of the same reproducibility issues as the compiled language packages investigated in the beginning of this section, they actually found that native extensions do not affect the distribution of irreproducible packages, and furthermore, do not expose issues other than the timestamps and file permissions that packages without native extensions suffer from.
They attribute this to the fact that for both, Python and NodeJS, the build systems building the native extensions have been optimized for reproducible builds already.
For Ruby no such optimized build system is used, so they speculate that due to the nature of native extensions as dedicated functions usually performing a small isolated task, the root causes for irreproducibility identified for compiled language ecosystems, which were usually found in larger software packages, do not play a role.~\cite{benedetti2025empirical}

\subsubsection{Artifact Reproduction}

A study of root causes for failing artifact reproduction has been performed by Goswami et al.~\cite{goswami2020investigating}.
As further described in \Cref{sec:state_of_the_art}, they tried reproducing all versions from the \num{1000} most depended upon packages.
However, they excluded various packages from their analysis, due to no longer being publicly available in the npm registry, not exposing a link to the source code repository, or not explicitly defining a build script, resulting in a sample of \num{3390} versions for \num{229} distinct packages.

They find that almost \qty{40}{\percent} of package versions are not reproducible.
This happened because the package was no longer publicly available in the npm registry, had no link to the source code
For the remaining \num{2898} packages, they investigated the differences in source code files, resulting in the following categories:
\begin{itemize}
    \item Different coding paradigms
    \item Different boolean expressions for the same logic
    \item Different number of statements or expressions
    \item Different variable names
    \item Different comments
    \item Different function or method order
    \item Different direct value assignments to variables
\end{itemize}
It is not further evaluated how much these differences in the code files cause differences in behaviour.~\cite{goswami2020investigating}

They derive three potential root causes:
Version relaxation for build dependencies, unclear transpiler versions and unclear minifier versions, resulting in their own build process potentially using different build tool versions than the distributed artifact.~\cite{goswami2020investigating}

We argue that the latter two are just specializations of version relaxation, thus version relaxation of build tools being the overarching root cause of these observed diffs.
Furthermore, we question the decision to exclude packages without a defined \texttt{build} script.
As previously mentioned, the \texttt{build} script is an optional, user-defined script that is not required to successfully build a NodeJS package.
At last, as noted in \Cref{sec:state_of_the_art}, the artifact they considered for their comparison is not the tarball that is actually distributed by the npm registry, but instead the individual files you get when unpacking this tarball, thus not taking any file metadata like timestamps or permission bits into account.
However, as shown by Benedetti et al.~\cite{benedetti2025empirical}, the \texttt{npm} package manager deterministically sets both of those, so it would likely not have come up as a problem anyway.

While not explicitly pursuing reproducible builds, the research by Vu et al.~\cite{vu2021py2src}, Gao et al.~\cite{gao2024pyradar} and Tsakpinis et al.~\cite{tsakpinis2024analyzing} aiming to retrieve source code repositories for given packages, highlight that getting the source code at all, as well as getting the correct revision for a given version, is an unsolved problem.
Even if registries were taking additional measures like enforcing links to source code or enforcing provenance information as specified by SLSA\footnote{\url{https://slsa.dev/spec/v1.0/}}, giving strong guarantees for source code availability is hard.
Not only does it depend on the willingness to publicly provide source code but also on external factors, like network availability to the source code hoster.

In summary, with regard to common problems for reproducible builds, we find that \emph{build timestamps} and \emph{file permissions} are identified as common causes for irreproducibility.
As already mentioned by Benedetti et al.~\cite{benedetti2025empirical}, an existing approach from the compiled language ecosystems, the \texttt{SOURCE\_DATE\_EPOCH} environment variable, is a viable solution to timestamp problems.
Regarding artifact reproduction, acquiring source code, usually in the form of a git repository, as well as finding the correct revision corresponding to a given artifact within this repository are two identified problems.
Additionally, unstable build dependencies resulting in other discrepancies are also found within both kinds of ecosystems.

In both cases, it has to be assumed that this is not a conclusive list.
While reproducible builds have been analysed across multiple ecosystems, there was only a single study investigating the isolated effects of the native build command, disregarding potential pre- and post-processing steps.
For example, any form of irreproducibility due to minification or transpilation, which is commonly used within NodeJS, is undetectable by the used methodology.
Studies on artifact reproduction only investigate single ecosystems for their respective focus areas, making it hard to generalize their findings across other ecosystems even within scripting languages.

\begin{rqanswer}
    \textbf{Response to RQ2:}
    The common challenges regarding reproducible builds are build timestamps and file permissions.
    Common challenges regarding artifact reproduction are the acquisition of source code repositories, as well as the identification of matching repository revisions to a given artifact.
    Due to the low number of studies for scripting language ecosystems, as well as the nature of the existing research, this is not assumed to be a conclusive list.
\end{rqanswer}

Regarding challenges unique to scripting language ecosystems, none are identified for reproducible builds.
For artifact reproduction, differences in source code files within artifacts are unique to scripting language ecosystems.
However, as noted above, their suspected root cause, unstable versions of build dependencies, are a known problem in compiled language ecosystems too.
Again, due to the small number of studies we assume this not to be a conclusive list, and that additional unique problems may be identified in the future, when a broader amount of ecosystems and build processes is analysed.

\begin{rqanswer}
    \textbf{Response to RQ3:}
    Regarding reproducible builds, no additional challenges are identified.
    Regarding artifact reproduction, additional challenges are various kinds of source code differences within released artifacts.
    Due to the low number of related studies within scripting language ecosystems, we cannot conclude that this is an exhaustive list.
\end{rqanswer}

\section{Discussion}\label{sec:discussion}

Reproducibility in compiled language ecosystems has been widely researched, and reproducibility in scripting language ecosystems is slowly catching up.
However, ambiguities within the context of the reproducible builds ecosystem lead to ambiguous and inconsistent use of the same terms within the literature, making it hard to compare existing results.

It is not clear what exactly is meant with build reproducibility, and what it is applied to.
The Reproducible Builds Project names it as a property of a \emph{build}, without further specifying what exactly a build is~\cite{reproducibleBuildsProject}, with other definitions not going into more detail either.
We propose \Cref{def:rep} as a more precise definition, derived from the most common way we have seen the term used in practice, matching the broad definition of the Reproducible Builds Project.
Additionally, current literature does not clearly differentiate between artifact reproduction and reproducible builds, as we have defined it.
Common definitions of these terms help unifying what exactly is subject of reproducibility research, and what methodology is used to test it.

However, even this definition may be improved.
It is undefined how exactly the build instructions are communicated to a building party.
For example, in the NodeJS ecosystem, a build process would at least require running the commands \texttt{npm install} to install build dependencies, followed by \texttt{npm pack} to create the package from the source code.
However, as noted before, there is a convention to also use the \texttt{npm build} command to execute build scripts, which is not explicitly endorsed by the ecosystem.
If the entry point, the command the building party has to run in order to run the communicated build script, is unclear, different building parties may execute different commands, resulting in different conclusions about reproducibility.
There appears to be the need for a universally agreed on entry point, at least within the scope of some context like an individual ecosystem, so that rebuilding parties know how to execute the build instructions.

Furthermore, we have also seen that the artifact to compare is not universally agreed on.
While most research uses the artifact as it is distributed by the respective registry, some perform post-processing like unpacking the artifact archive and then comparing the contained files.
An argument may be made that archive metadata is irrelevant to the behaviour of a certain artifact, since package managers often unpack them anyway.
However, different file names or permissions might lead to different behaviour of the artifact during execution, but would not be caught by content only comparisons.
Furthermore, especially when considering a malicious actor, the process of archive extraction may have unexpected side effects, which may not be noticed when comparing the apparent contents.
For example, \texttt{tar} extraction may place files outside the current working directory without proper safeguards, possibly resultin in malicious files in places unnoticed by a user.
We would therefore argue that the artifact to compare shall always be the artifact as it is distributed by the respective registry.

Doing such preprocessing of the distributed artifact may also be considered using a level 2 equivalence function instead of a level 1 equivalence function, leaving out the seemingly insignificant metadata.
Inconsistent use of these comparison functions are also present across the industry, as for example the Fedora Project uses a level 2 equivalence function, disregarding embedded signatures~\cite{fedoraReproducibleBuilds}, while the Reproducible Builds Project specifies the use of a level 1 equivalence function~\cite{reproducibleBuildsProject}.
However, at least within the existing research, the term \enquote{reproducible builds} is always used in terms of level 1 equivalence, and any work deviating from this has coined their own term, like \enquote{verifiable builds}~\cite{carnavalet2014challenges}.

Furthermore, neither for reproducible builds, nor for artifact reproduction, there is a clear definition on when something counts as reproducible.
Regarding reproducible builds, the most robust approach appears to be adversarial rebuilding (c.f. \Cref{sec:background}).
However, there is no universal agreement on which external variables to change in what ways in order to label something as reproducible.
Thus it could happen that builds are labelled as reproducible when they actually are not reproducible, if the irreproducibility causing external factors were not modified in the right way.
For artifact reproduction, the case is less clear due to ambiguities like the aforementioned missing entry point.
If building parties assume different entry points for the build script, they might come to different conclusions regarding the artifact's reproducibility.

This lack of standardization leads to difficulties when comparing research results, as they cause differences in methodologies in a way that results which appear to mean the same actually do not.
Two studies claiming to analyse reproducibility of NodeJS packages, one concluding \qty{100}{\percent} reproducibility and the other claiming \qty{10}{\percent} reproducibility do not contradict each other, if one study performed registry artifact reproduction and the other compared adversarially rebuilt artifacts.

\section{Future Research Directions}\label{sec:future-research-directions}
The provided review of literature allows us to identify multiple future research directions to advance the knowledge concerning reproducible builds in scripting language ecosystems.

With regard to the points of ambiguity mentioned in \Cref{sec:discussion}, procedures should be developed that allow for standardized research within and across ecosystems.
While ambiguous naming is also a concern, research would mostly be required to create generic, standardized processes for both, registry artifact reproduction and adversarial rebuilding.
A standardized format for communicating the build under test, as well as standard entry points for artifact reproduction have to be developed.
Technologies that recently gained popularity, like provenance information with attestations or software bills of material might be promising candidates to contain parts of this information.

Regarding adversarial rebuilding, a set of environment variations shall be introduced which all builds under test should be subject to.
If the artifacts stay the same under these variations, they can be labelled reproducible.
This would help to standardize future reproducibility research, leading to more streamlined methodologies and more comparable results.

A taxonomy of root causes for irreproducibility regarding both artifact reproductions and reproducible builds, including a frequency and impact analysis similar to the work Bajaj et al.~\cite{bajaj2023unreproducible} have done for compiled language ecosystems, would be helpful to gain insights into categories of issues, and the ability to focus on high impact root causes.
The work by Benedetti et al.~\cite{benedetti2025empirical} lays good groundwork regarding reproducible builds, and could be expanded to include more pre-processing steps, like executing the \texttt{build} script for NodeJS packages.

Finding root causes for discrepancies between artifact reproduction and reproducible builds results would also be helpful.
Ideally, the registry artifact corresponding to any build that is regarded reproducible should also be able to be reproduced.
However, we can see that this is not the case, for example comparing the results shown by Benedetti et al.~\cite{benedetti2025empirical} and Goswami et al.~\cite{goswami2020investigating}.
The reason for this probably lies in arbitrary pre- and post-processing steps executed by the project maintainer when creating the artifact that is distributed by the registry, but are not obvious to a rebuilding party.
Finding out if there are typical steps in the respective ecosystems, like the aforementioned \texttt{build} script for NodeJS packages, could help to standardize build instructions rebuilding parties shall perform, and while this standardization process is ongoing, could also assist in closing the gap between artifact reproduction and reproducible builds research.

Once more reasons for irreproducibility in scripting language ecosystems have been established, more research into fixing these problems can be performed.
Again, the work by Benedetti et al.~\cite{benedetti2025empirical} shows a good example, transferring the use of the \texttt{SOURCE\_DATE\_EPOCH} environment variable from compiled language ecosystems to scripting language ecosystems, and proposing patches to package managers that fix non-deterministic file permission bits.

\section{Conclusion}\label{sec:conclusion}

In this work we have analysed literature regarding reproducibility research, with a focus on scripting language ecosystems.
We have shown that in general, research in this field is still sparse, but the concentration of publications in recent years indicates that it is gaining popularity.
In contrast to compiled language ecosystems, the literature focuses more on reproducing distributed artifacts, than proving the reproducibility of the build process itself.

The identified reasons for irreproducibility so far only consist of a small subset of root causes identified for compiled language ecosystems, in addition to differences in source code files of the distributed artifact, which are unique to scripting language ecosystems by nature.
However, due to the little research done so far and individual studies focusing on single ecosystems, we assume that more root causes may be discovered in the future.

Furthermore, we pointed out problems with comparability of research results, due to ambiguities in existing definitions of reproducibility, and resulting inconsistencies in their interpretation in the literature.
We proposed more rigid notions of reproducible build and artifact reproduction, and advocate for more standardization in evaluation processes in the future.

\def\UrlBreaks{\do\/\do-}
\bibliographystyle{splncs04}
\bibliography{bibliography}

\end{document}